\documentclass[dvips,12pt]{article}

\usepackage{amsmath,amsthm,amssymb,epsf,epsfig,a4wide}

\theoremstyle{definition}
\newtheorem{theorem}{Theorem}

\newtheorem{lemma}[theorem]{Lemma}
\newtheorem{proposition}[theorem]{Proposition}
\newtheorem{definition}[theorem]{Definition}

\begin{document}

\begin{centering}

\vspace{0in}

{\Large {\bf Intersecting hypersurfaces
in dimensionally continued topological density gravitation}}

\vspace{0.4in}

{ \bf Elias Gravanis, \; Steven Willison} \\
\vspace{0.2in}
Department of Physics, King's College London,\\
Strand, London WC2R 2LS, United Kingdom.

\vspace{0.2in}
{\bf Abstract}

\end{centering}

\vspace{0.2in}

{\small We consider intersecting hypersurfaces in curved 
spacetime with gravity governed by a class 
of actions which are topological invariants in lower 
dimensionality. Along with the Chern-Simons 
boundary terms there is a sequence of intersection 
terms that should be added in the action 
functional for a well defined variational principle. 
We construct them in the case of Characteristic 
Classes, obtaining relations which have a general 
topological meaning. Applying them on a manifold 
with a discontinuous connection 1-form we obtain 
the gravity action functional of the system and show 
that the junction conditions can be found in a simple 
algebraic way. At the sequence of intersections 
there are localised independent energy tensors, 
constrained only by energy conservation. We work out 
explicitly the simplest non trivial case.}
\vspace{0.3in}

{\bf Introduction}

\vspace{0.1in}

General relativity can be generalised to a manifold with 
boundary. The inclusion of a certain boundary term 
(Gibbons-Hawking) makes the action principle well defined 
on the boundary. Also, singular hypersurfaces of matter\cite{Israel} 
can be incorporated into a manifold with piece-wise differentiable
metric\cite{noteC1}. 
We will see that these are part of the general properties of actions built 
out of dimensionally continued topological invariants. The Einstein-Hilbert 
action of General Relativity is the dimensionally continued form of the two 
dimensional Euler Characteristic. A linear combination of terms  which are 
dimensionally continued Euler densities in arbitrary dimensions is known
variously as Lovelock or Lanczos-Lovelock or Gauss-Bonnet gravity. It has 
been studied 
extensively$\cite{Lovelock,Zumino,Aragone,Deruelle}$ and the boundary action 
has been constructed\cite{Myers,Muller-Hoissen,Verwimp}.

An interesting problem in gravity is the study of collisions of shells of 
matter\cite{Dray:yt,Langlois:2001uq,Berezin-02}. 
Brane-world models of matter on the intersection of co-dimension 1 branes 
were studied and it was found that the Gauss-Bonnet term was needed in order to get 
a tension on the intersection in a natural way\cite{Lee}. The Gauss-Bonnet term is the 
dimensionally continued 4-dimensional Euler density.
We address this problem of intersections and collisions of co-dimension 1
hypersurfaces in a more
generalised way, motivated by the properties of the topological invariants.

A topological invariant `action' contains no local degrees of freedom. The 
only information it
encodes is topological. As such it is independent of the local form of the 
metric. We consider actions which are topological in a certain dimensionality 
and then generalise to higher dimensions. These actions have the property that 
the independent infinitesimal variation of the action with respect to the 
connection is a total derivative. 

We consider a smooth manifold with embedded arbitrarily intersecting 
hypersurfaces of singular matter. We can view a hypersurface as the shared 
boundary of two adjacent regions. The gravity of localised matter can be 
described by a boundary action. As in \cite{Lee}, we allow for the possibility of 
matter being localised on the surfaces of intersection also. The spacetime is 
divided up into polyhedral regions bounded by piece-wise smooth hypersurfaces- 
like a matrix of cells. We show that this situation is compatible with any theory 
of gravity based on a dimensionally continued topological invariant.

We exploit the topological nature of the theory to write the action in terms of 
different connections in different regions. This generates our surface actions 
remarkably simply. We can derive the Israel junction conditions and the junction 
conditions for any intersections in a purely algebraic way.

In section 1 we review basic material on topological densities, introducing 
Characteristic Classes on a manifold with boundary. In section 2.1 we derive the 
intersection forms generalising Chern-Simons forms and in section 2.2 we construct 
the action functional of gravity in the presence of intersecting
hypersurfaces employing the properties of the intersection forms. The topological 
theory is dealt with in
section 2.2.1 and the dimensionally continued case of interest is dealt with in 
section 2.2.2. In section 3 we work out a simple example along with the energy 
exchange relations in that case.

\section{Topological densities and gravitation}

Let M be a manifold with a Riemannian or Lorentzian metric $g$ and a 
Levi-Civita connection. Let $\omega$ be the connection 1-form and $\Omega$ 
the curvature form. For ${\rm dim}M=2n$ consider
the integral
\begin{equation}
\int_M f(\Omega,..\Omega), \qquad  f(\Omega,..,\Omega)=\Omega^{a_1a_2} \wedge ...
\wedge \Omega^{a_{2n-1}a_{2n}} \epsilon_{a_1...a_{2n}}
\end{equation}
where $\epsilon_{...}$ is the fully anti-symmetric symbol and $\epsilon_{1..2n}=+1$ 
and the 
integral is assumed to exist. The frame $E$ is ortho-normal in the sense 
$g(E^a,E^b)=\delta^{ab}$ 
in the Riemannian case and $g(E^a,E^b)=\eta^{ab}=diag{(-1,1..1)}$ in the 
Lorentzian case. 

When $g$ is Riemannian and M is compact and oriented $f(\Omega,..\Omega)$ 
represents the Euler class. The integral  over M, normalised properly, gives 
the Euler number of M, according to the renowned Gauss-Bonnet-Chern 
Theorem$\cite{Chern}$. There are more general (and precise) definitions than the 
one we give here. The details are in the textbooks$\cite{Koba,mathbook2,Eguchi}$ 
but we only need to point out the similarity and make intelligible borrowing of 
tools from global differential geometry. $f(\Omega,..\Omega)$ will be called the 
Euler density. In general an invariant whose integral
over M gives a topological invariant of M will be called topological density.

Let us now repeat the Chern-Weil construction and show that under a continuous 
change of the connection, $\omega \to \omega'$, $f(\Omega,..\Omega)$ changes 
by an exact form. In fact we are only going to need $f$ to be invariant, 
symmetric and multi-linear. These general properties are provided by the 
invariant polynomials and lead to the so-called 
Characteristic Classes, of which the Euler class is an example. 
The following applies globally on the principal bundle but it is sufficient 
for our purpose to work on the manifold.

Define
\begin{equation*}
\omega_t=t \omega+(1-t)\omega'
\end{equation*}
Call
\begin{equation*}
\theta=\omega-\omega'
\end{equation*}
and note that
\begin{equation*}
\theta=\frac{d}{dt} \omega_t
\end{equation*}
and for the curvature associated with $\omega_t$
\begin{equation}
\Omega_t=d\omega_t+\omega_t \wedge \omega_t
\end{equation}
that
\begin{equation}
\frac{d}{dt} \Omega_t=D_t \theta
\end{equation}
where $D_t$ is the covariant derivative associated with $\omega_t$. Then
\begin{eqnarray} \label{derivation1}
&& f(\Omega,..,\Omega)-f(\Omega',..,\Omega')= \int_0^1 dt
\frac{d}{dt} f(\Omega_t,..,\Omega_t)= n \int_0^1 dt f( d
\Omega_t/dt, \Omega_t, ...\Omega_t)= \\ \nonumber && =n \int_0^1
dt f(D_t \theta, \Omega_t, ...\Omega_t)= n \int_0^1 dt  \ d f(
\theta, \Omega_t, ...\Omega_t)
\end{eqnarray}
where symmetry and multi-linearity of $f$ have been used, as well as $D_t\Omega_t=0$.

If we define
\begin{equation}
{\cal L}(\omega)=f(\Omega,..\Omega)
\end{equation}
and
\begin{equation}
{\cal L}(\omega,\omega')=-n \int_0^1 dt f(\omega-\omega',
\Omega_t, ...\Omega_t)
\end{equation}
we can write
\begin{equation} \label{basic}
{\cal L}(\omega)={\cal L}(\omega')-d{\cal L}(\omega,\omega')
\end{equation}
Now, assume that, for example, M is non-compact and without a boundary. If 
${\cal L}(\omega,\omega')$ vanishes fast enough asymptotically
\begin{equation}
\int_M {\cal L}(\omega)
\end{equation}
(assumed to exist) does not depend on $\omega$. It is this property that makes 
${\cal L}(\omega)$ so useful when, with a little modification, it is used as a 
Lagrangian for gravity for ${\rm dim}M>2n$.

Define the (d-r)-form (which is a natural (d-r)-dimensional volume element):
\begin{equation}
e_{a_1a_2...a_r}=\frac{1}{(d-r)!} \ \epsilon_{a_1a_2...a_d}
E^{a_{r+1}} \wedge .. \wedge E^{a_d}.
\end{equation}
The associated dimensionally continued Euler density for $d>2n$ is
\begin{equation} \label{gravaction}
{\cal L}_g(\omega,e)=f(\Omega,..,\Omega,e)=\Omega^{a_1b_1} \wedge
\Omega^{a_2b_2} \wedge .. \wedge \Omega^{a_n b_n} \wedge e_{a_1b_1
a_2 b_2...a_n b_n}
\end{equation}
which is also an invariant.

Then, the Euler-Lagrange variation with respect to $\omega$ in
\begin{equation}
\int_M {\cal L}_g (\omega,e),
\end{equation}
noting that $\delta \Omega= D(\delta \omega)$, vanishes by the Bianch 
identity and the assumed zero torsion condition 
$D E^a=0$\cite{Pontryagin_note}. The equations of motion are obtained simply 
by the Euler-Lagrange variation of the frame, applying the formula
\begin{equation} \label{varvol}
\delta e_{a_1..a_r}=\delta E^{a_{r+1}} \wedge e_{a_1..a_ra_{r+1}}
\end{equation}
in a purely algebraic way.

In the next section we find how the action (\ref{gravaction}) is re-expressed 
in the presence of hypersurfaces and their intersections, by generalising 
(\ref{basic}) appropriately, and show that
the equations of motion (junction conditions) are still obtained from the 
mere variation of the frame.

\section{Topological densities on manifolds containing intersecting
hypersurfaces}\label{topden}

A hypersurface is understood as a smooth co-dimension 1 subspace of the manifold 
where the connection form exhibits discontinuity or as a (higher co-dimension) 
intersection of such discontinuities.

Integrating ${\cal L}(\omega)$ over the manifold, when $\omega$ is the discontinuous 
connection form, one has to add a Chern-Simons term integrated over the discontinuity 
for the final result to have well defined variations with respect to $\omega$ (and to 
be diffeomorphism invariant).
If discontinuities intersect, in all possible ways, one should, in general, add 
appropriate generalisations of the Chern-Simons forms integrated over the 
intersections.

A discontinuity can be thought of as the common boundary of two d-dimensional (bulk) 
regions. Intersection of discontinuities can be thought of as common subspaces of 
the (not smooth) boundaries of a larger number of bulk regions. It is also helpful 
to think of them as singular overlaps (at the boundaries) or intersections of two 
or more bulk regions. 

With this in mind, we first find generalisations of the Chern-Simons forms.

\subsection{From boundary to intersection action terms}\label{biat}

Given an invariant polynomial, we found in the previous section a
relation of the form of (\ref{basic}) by interpolating between the
given connection $\omega$ and an arbitrary one $\omega'$. We can
continue by interpolating between the latter and a new connection.
In general, let us define the p-parameter family of connections,
interpolating between p+1 connections, $\omega^1,..,\omega^{p+1}$,
\begin{equation}
\omega_p=\omega_{t_1..t_p}=\omega^{1}-(1-t_1)
\theta^{1}-...-(1-t_1)..(1-t_p) \theta^{p}
\end{equation}
where
\begin{equation}
\theta^{r}=\omega^{r}-\omega^{r+1}, \quad r=1,..,p .
\end{equation}
Note: For the purposes of section \ref{biat} and equations
\ref{palatini} to \ref{eqm} only, the subscript p refers to a
function of $t_1,...t_p$. 

Define
\begin{equation}
\frac{\partial}{\partial t_q}\omega_{p}= \frac{\partial}{\partial
t_q}\omega_{t_1..t_p}= \sum^p_{r \geq q} (1-t_1)..
\widehat{(1-t_q)}...(1-t_r) \theta^{r}=
\theta^{q}_{t_1...\widehat{t_q}...t_p}=\theta^q_p .
\end{equation}
where the hat means that the index is omitted. Note that
\begin{equation} \label{omission}
\omega_{t_1..t_p}|_{t_r=0}=\omega_{t_1..t_{r-1}t_{r+1}..t_p}
\end{equation}
setting the connection $\omega^r=0$. This will be useful below.
Let $\Omega_p$ be the p-parameter curvature 2-form associated with
$\omega_p$. Then
\begin{equation} \label{id1}
\frac{\partial}{\partial t_q}\Omega_{p}= \frac{\partial}{\partial
t_q}\Omega_{t_1..t_p}= D_{t_1..t_p}
\theta^{q}_{t_1...\widehat{t_q}...t_p}=D_p\theta^q_p.
\end{equation}
There is also a ``Bianchi identity" for $\Omega_p$
\begin{gather}
D_p\Omega_p = 0
\end{gather}
$D_p$ is the covariant derivative associated with $\omega_p$.

\begin{proposition}
We introduce a (p+1)-point term with p+1 connection entries. We now show that 
the (p+1)-point generalization of the 2-point Chern-Simons term takes the form
\begin{equation} \label{genl}
{\cal L}(\omega^1,..,\omega^{p+1})= \eta_p \frac{n!}{(n-p)!}
\int_0^1 dt_1..dt_p \
f(\theta^1_p,\theta^2_p,..\theta^p_p,\Omega_p,..\Omega_p)
\end{equation}
where $\eta_p=(-1)^{\frac{p(p+1)}{2}}$. These terms obey the following rule:
\begin{equation}\label{comp1}
\sum_{s=1}^{p+1} (-1)^{s-p-1} {\cal
L}(\omega^1,..,\widehat{\omega^s},..,\omega^{p+1},\omega^{p+2})=
{\cal L}(\omega^1,..,\omega^{p+1})+d{\cal
L}(\omega^1,..,\omega^{p+1},\omega^{p+2}).
\end{equation}
\\\\
{\bf Proof:} If we define
\begin{equation}
\omega^{p+1}_{t_{p+1}}=t_{p+1} \omega^{p+1} + (1-t_{p+1})
\omega^{p+2}
\end{equation}
then
\begin{equation*}
{\cal L}(\omega^1,..,\omega^{p+1}_{t_{p+1}})= \eta_p
\frac{n!}{(n-p)!} \int_0^1 dt_1..dt_p \
f(\theta^1_{p+1},\theta^2_{p+1},..\theta^p_{p+1},\Omega_{p+1},..\Omega_{p+1})
\end{equation*}

We have
\begin{eqnarray*}
&& {\cal L}(\omega^1,..,\omega^p,\omega^{p+1})- {\cal
L}(\omega^1,..,\omega^p,\omega^{p+2})= \int_0^1 dt_{p+1}
\frac{\partial}{\partial t_{p+1}} {\cal
L}(\omega^1,..,\omega^p,\omega^{p+1}_{t_{p+1}}) \\ && = \eta_p
\frac{n!}{(n-p)!} \int_0^1 dt_1..dt_p
dt_{p+1} \ \frac{\partial}{\partial t_{p+1}} \
f(\theta^1_{p+1},\theta^2_{p+1},..\theta^p_{p+1},\Omega_{p+1},..\Omega_{p+1})
\nonumber
\end{eqnarray*}
From the multi-linearity of the invariant polynomial $f$ we have
\begin{eqnarray*}
&& \frac{\partial}{\partial t_{p+1}}
f(\theta^1_{p+1},\theta^2_{p+1},..\theta^p_{p+1},\Omega_{p+1},..\Omega_{p+1})=
\\ &&
\sum_{r=1}^p f(\theta^1_{p+1},..,\frac{\partial}{\partial t_{p+1}}
\theta^r_{p+1},..,\theta^p_{p+1},\Omega_{p+1},..\Omega_{p+1})+
\\\nonumber && (n-p)
f(\theta^1_{p+1},\theta^2_{p+1},..\theta^p_{p+1},
\frac{\partial}{\partial t_{p+1}}\Omega_{p+1},..\Omega_{p+1})
\nonumber
\end{eqnarray*}
Using (\ref{id1}) we can write the last term as
\begin{eqnarray*}
&& (n-p) (-1)^p \
df(\theta^1_{p+1},\theta^2_{p+1},..\theta^p_{p+1},\theta^{p+1}_{p+1},
\Omega_{p+1},..\Omega_{p+1})  \\ && -(n-p) \sum_{s=1}^p
(-1)^{p+s-1} f(\theta^1_p,..,D_{p+1}
\theta^s_{p+1},..,\theta^{p+1}_{p+1}, \Omega_{p+1},..\Omega_{p+1})
\nonumber
\end{eqnarray*}
and using again (\ref{id1}) in the last term we obtain
\begin{eqnarray*}
&&
-\sum_{s=1}^p (-1)^{p+s-1} \frac{\partial}{\partial t_s}
f(\theta^1_{p+1},..,\widehat{\theta^s_{p+1}},..,\theta^{p+1}_{p+1},
\Omega_{p+1},..\Omega_{p+1}) \\ && +\sum_{s=1}^p (-1)^{p+s-1}
\sum_{r=1, \neq s}^{p+1}
f(\theta^1_{p+1},..,\frac{\partial}{\partial t_s}
\theta^r_{p+1},..
\widehat{\theta^s_{p+1}},..,\theta^{p}_{p+1},\theta^{p+1}_{p+1},
\Omega_{p+1},..\Omega_{p+1}) \nonumber
\end{eqnarray*}
In all
\begin{eqnarray*}
&& (n-p) (-1)^p
df(\theta^1_{p+1},\theta^2_{p+1},..\theta^p_{p+1},\theta^{p+1}_{p+1},
\Omega_{p+1},..\Omega_{p+1})  \\ && \nonumber
-\sum_{s=1}^p (-1)^{p+s-1}
\frac{\partial}{\partial t_s}
f(\theta^1_{p+1},..,\widehat{\theta^s_{p+1}},..,\theta^{p+1}_{p+1},
\Omega_{p+1},..,\Omega_{p+1}) \\\nonumber &&
+\sum_{s=1}^{p+1}(-1)^{p+s-1} \sum_{r=1, \neq s}^{p+1}
f(\theta^1_{p+1},.., \frac{\partial}{\partial t_s}
\theta^r_{p+1},..,
\widehat{\theta^s_{p+1}},..,\theta^{p}_{p+1},\theta^{p+1}_{p+1},
\Omega_{p+1},..,\Omega_{p+1}) \nonumber
\end{eqnarray*}
Note now that
\begin{equation}
\frac{\partial}{\partial t_s} \theta^r_{p+1}=
\frac{\partial}{\partial t_s} \frac{\partial}{\partial t_r}
\omega_{p+1}= \frac{\partial}{\partial t_r} \theta^s_{p+1}
\end{equation}
then in the last term, if we split the sum into $r<s$ and $r>s$, changing
variables $ r \leftrightarrow s$ in the latter and using this
identity we see that the term vanishes. We have shown then that
(p+1)-point ${\cal L}_{p+1}$ defined in (\ref{genl}) obeys a rule
\begin{equation*}
\sum_{s=1}^{p+1} (-1)^{s-p-1} {\cal
L}(\omega^1,..,\widehat{\omega^s},..,\omega^{p+1},\omega^{p+2})=
{\cal L}(\omega^1,..,\omega^{p+1})+d{\cal
L}(\omega^1,..,\omega^{p+1},\omega^{p+2})
\end{equation*}
The relation (\ref{omission}) has been used.$\Box$
\end{proposition}

It is not hard to show that ${\cal L}$ is fully anti-symmetric in
its entries, so we can write the above in the form
\begin{equation} \label{ncomprule}
\sum_{s=1}^{p+1}  {\cal
L}(\omega^1,..,\omega^{s-1},\omega',\omega^{s+1},..,\omega^{p+1})=
{\cal L}(\omega^1,..,\omega^{p+1})+d{\cal
L}(\omega^1,..,\omega^{p+1},\omega')
\end{equation}
where $\omega'$ is arbitrary.

As $\theta^r$ is a 1-form we have
$f(..,\theta^r,..,\theta^r,..,\Omega_{p},..\Omega_{p})=0$ and we
can write (\ref{genl}) explicitly in terms of
$\theta^r=\omega^r-\omega^{r+1}$, $r=1..p\ $ in the form
\begin{eqnarray} \label{genl2}
&& {\cal L}(\omega^1,..,\omega^{p+1})=  \int_0^1 dt_1..dt_p \
\zeta_p \  f(\theta^1,\theta^2,..\theta^p,\Omega_p,..,\Omega_p),
\\ &&  \label{zeta}
\zeta_p=(-1)^{\frac{p(p+1)}{2}} \frac{n!}{(n-p)!} \ \
\prod_{r=1}^{p-1}(1-t_r)^{p-r}.
\end{eqnarray}
\\

Let us show that ${\cal L}_p$'s, constructed from Characteristic 
Classes, are invariant under
local Lorentz transformations. The
connections transform as
\begin{equation}
\omega^r_{(g)}=g^{-1} \omega^r g+ g^{-1} dg
\end{equation}
for all $r=1,..,p+1$, where $g$ belongs to the
adjoint representation of SO(d-1,1). Then, $\theta^r_{(g)}=g^{-1} \theta^r g$ 
and $\Omega_{p(g)}=g^{-1} \Omega_p g$, so
\cite{CSboundary}
\begin{equation}
\label{ginv}
{\cal L}(\omega^1_{(g)},..,\omega^{p+1}_{(g)})={\cal 
L}(\omega^1,..,\omega^{p+1}).
\end{equation}
\\

In fact, one can derive (\ref{ncomprule}) without reference to the invariant
polynomial , by use of the Poincare lemma and the following observation (inspired by the
form of (\ref{ncomprule})). If $f(x_1,..x_n)$ is an anti-symmetric function  of n variables
and
\begin{equation}
Af(x_1,..x_p,x_{p+1})=f(x_1,..x_p)-\sum_{i=1}^p
f(x_1,..,x_{i-1},x_{p+1},x_{i+1},..x_p)
\end{equation}
(antisymmetrising over n+1 variables) then $AAf(x_1,..,x_{p+2})=0$.
The proof is trivial.

We can now show (\ref{ncomprule}) by induction assuming only that ${\cal 
L}(\omega)$ obeys (\ref{basic}). I.e. it is true for the p=0 case. Assume
(\ref{ncomprule}) for p=k-1 (let us use the symbol ${\cal L}_k (\omega^1..\omega^k)$
for the intersection forms in this proof)
\begin{equation}
A{\cal L}_k (\omega^1..\omega^{k+1})=
-d{\cal L}_{k+1}(\omega^1..\omega^{k+1})
\end{equation}
Then $dA{\cal L}_{k+1} (\omega^1..\omega^{k+2})=0$. By the Poincare lemma
we have that there exists an invariant form, locally, such that
\begin{gather}
A{\cal L}_{k+1}(\omega^1..\omega^{k+2})=
-d{\cal L}_{k+2}(\omega^1..\omega^{k+2})
\end{gather}
which completes the induction.  
(\ref{genl}) is {\it a} solution of the general relation (\ref{ncomprule}).

 There is similarity between our composition rule and
Stora-Zumino descent equations $\cite{Ginsparg}$.
  The reason is the existence in both cases of a nilpotent operator, $A$ in
our case and the fermionic BRST operator there, which commutes and
anticommutes respectively with the derivative operator $d$.

\subsection{Manifolds with discontinuous connection 1-form}

We now construct the action functional of gravity on a manifold
containing intersecting surfaces. It will also enable us to draw
conclusions for arbitrary intersections of hypersurfaces for a
general dimensionally continued topological density.

\subsubsection{Topological density}
If the functional $\int_M {\cal L}$ is independent of the $C^0$
metric of the manifold M, then it can be evaluated using a continuous
connection as well as a connection that is discontinuous at some subspaces
(namely there are hypersurfaces involved). That is, the result will be the
same. We use this formal equivalence to give a meaning to $\int_M {\cal L}(\omega)$ when
$\omega$ is discontinuous.

Let us start with the case of a topological density ${\cal L}(\omega_0)$
of a continuous connection $\omega_0$ integrated
over M which contains a single hypersurface. Label 1 and 2 the regions
of M separated by the hypersurface. Introduce two
connections, $\omega_1$ and $\omega_2$ which are smooth in the
regions 1 and 2 respectively. We now write
\begin{gather} \label{1}
\int_M {\cal L}(\omega_0) = \int_{1}{\cal L}(\omega_1)+d{\cal
L}(\omega_1,\omega_0) +\int_{2}{\cal L}(\omega_2)+d{\cal
L}(\omega_2,\omega_0)
\end{gather}
Label the surface, oriented with respect to region 1, with 12.
(Formally $\int_{12} = -\int_{21}$).
\begin{align}\nonumber  \label{2}
\int_M {\cal L}(\omega_0) =& \int_{1}{\cal L}(\omega_1)
+\int_{2}{\cal L}(\omega_2)+ \int_{12}{\cal
L}(\omega_1,\omega_0)-{\cal L}(\omega_2,\omega_0)
\\=&\int_{1}{\cal L}(\omega_1)
+\int_{2}{\cal L}(\omega_2)+ \int_{12}{\cal
L}(\omega_1,\omega_2)+d{\cal L}(\omega_1,\omega_2,\omega_0)
\end{align}
That is, for a smooth surface the r.h.s. is independent of
$\omega_0$.

Consider now a sequence of co-dimension $p=1,2,3..h$ hyper-surfaces which
are intersections of $p+1=2,3..h+1$ bulk regions respectively.
We will use the terms intersection and hyper-surface alternatively. A
co-dimension p hyper-surface is labelled by $i_0..i_p$ where $i_0,..,i_p$
are the labels of the bulk regions which intersect there. We call this
configuration a simplicial intersection.

\begin{figure}\label{3inter}
\begin{center}\mbox{\epsfig{file=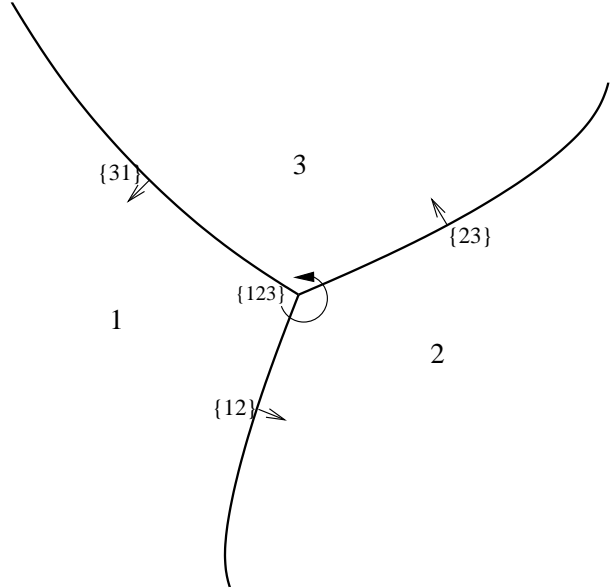, width=8cm}}
\caption{{\small The simplicial intersection of co-dimension 2 (h=2). 
The totally
antisymmetric symbol \{123\} specifies the intersection including the
orientation}}
\end{center}
\end{figure}

\begin{figure}\label{4inter}
\begin{center}\mbox{\epsfig{file=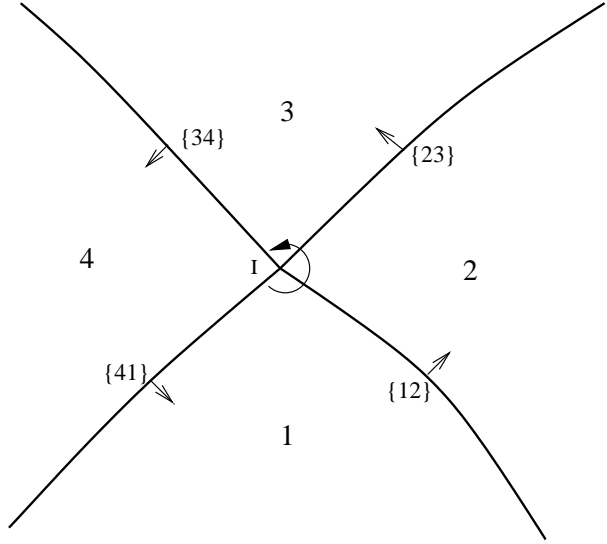, width=8cm}}
\caption{{\small A non-simplicial intersection of co-dimension 2 (h=2). The 
intersection, including the orientation, would not be properly represented
by the totally anti-symmetric symbol $\{1234\}$}}
\end{center}
\end{figure}

We take the example $h=2$ (fig. \ref{3inter}),
where the intersections are
$\{12\}$, $\{13\}$, $\{23\}$, $\{123\}$. An exact form integrated over 
$\{12\}$ will 
contribute at $\{123\}$
the opposite that when integrated over $\{21\}$, that is, for the latter
integration the intersection can be labelled by $-123=213$, if
we assume anti-symmetry of the label. The arrows of positive
orientations in fig. 1 tell us that a fully anti-symmetric symbol
$\{123\}$ will adequately describe the orientations of the intersection
$123$. This is in contrast to the non-simplicial intersection (fig. \ref{4inter}).
\begin{definition}[for a simplicial intersection]
$\{i_0...i_p\}$ is the set $\overline{i_0}\cap
\cdots \cap \overline{i_p}$
where $\overline{i_r}$ is the closure of the open set $i_r$ (a bulk region).
$\overline{i_r}$ overlap such that $\partial i_r = \sum_{s=0,\neq r}^h
 \overline{i_r}\cap\overline{i_s}$
and $i_r \cap i_s = \emptyset$ for all $s\neq r$. 
This formalises our definitions at the beginning of Section \ref{topden}.
By $\partial (\overline{A}\cap \overline{B}) = (\partial \overline{A}\cap 
\overline{B})\cup  (\overline{A}\cap \partial\overline{B})$, for A, B open sets,
we can write: 
\begin{gather}\label{defthesets}
\partial \{i_0...i_p\} = \sum_{i_{p+1}} \{i_0...i_{p+1}\}.
\end{gather}
Full anti-symmetry of the symbol $\{i_0...i_p\}$ keeps track of the orientations
properly in (\ref{defthesets}). As a check:
\begin{gather*}
\partial^2 \{i_0...i_p\} = \sum_{i_{p+1},i_{p+2}} \{i_0...i_{p+1}i_{p+2}\}=0.
\end{gather*}
\end{definition}

\begin{lemma}\label{lem3}
When all intersections are simplicial intersections, with no localised
curvature,
the contribution from each intersection $\{i_1...i_k\}$ is
\begin{gather}
\int_{\{i_1...i_k\}} {\cal L}(\omega_{i_1},...,\omega_{i_k})
\end{gather}
up to a boundary term on $\partial M$.

By ``no localised curvature," it is meant that
the distributional part of the Riemann curvature tensor must have its 
support only on
co-dimension 1 hypersurfaces and not on lower dimensional intersections.
This is an important condition. We require this 
in order to have a well defined ortho-normal frame at the intersections.
\\\\
{\bf Proof:}
Assume, for  $l < h$, we can write
\begin{gather}
\int_M {\cal L}(\omega_0)= \sum^{l-1}_{k=1} 
\frac{1}{k!} \sum_{i_1..i_k} \int_{\{i_1...i_k\}}
{\cal L}(\omega_{i_1}..\omega_{i_k}) +
\frac{1}{l!} \sum_{i_1..i_l}\int_{\{i_1...i_l\}}
{\cal L}(\omega_{i_1}..\omega_{i_l}) +
d{\cal L}(\omega_{i_1}..\omega_{i_l},\omega_0)  . 
\end{gather}
We have already seen that this is true for $l=1$ and $l=2$. 
The exact form gives
\begin{equation*}
\frac{1}{l!} \sum_{i_1..i_li_{l+1}} \int_{\{i_1...i_li_{l+1}\}}
{\cal L}(\omega_{i_1}..\omega_{i_l},\omega_0)
\end{equation*}
$+$ a term on $\partial M$.
From the anti-symmetry of $\{i_1..i_li_{l+1}\}$ and of ${\cal
L}$ we have
\begin{equation*}
\frac{1}{l!} \sum_{i_1..i_{l+1}} \int_{\{i_1...i_li_{l+1}\}}
\frac{1}{l+1} \sum_{r=1}^{l+1} {\cal
L}(\omega_{i_1}..\omega_{i_{r-1}},\omega_0,\omega_{i_{r+1}}..
\omega_{i_{l+1}})
\end{equation*}
Applying the composition rule we get:
\begin{align}
\int_M {\cal L}(\omega_0)= & \sum^{l}_{k=1} \frac{1}{k!} 
\sum_{i_1..i_k} \int_{\{i_1...i_k\}}
{\cal L}(\omega_{i_1}..\omega_{i_k}) \nonumber\\ & +
\frac{1}{(l+1)!}\sum_{i_1..i_{l+1}} \int_{\{i_1...i_{l+1}\}}
{\cal L}(\omega_{i_1}..\omega_{i_{l+1}})  
+d{\cal L}(\omega_{i_1}..\omega_{i_{l+1}},\omega_0)  
\end{align}
Finally we note that the total derivative term
on the highest co-dimension intersections (order h), can only contribute 
to $\partial M$.
So by induction we have proved the Lemma.

Note that apart from our composition formula we have used only Stokes' 
theorem, which is valid on
a topologically non-trivial manifold M assuming a partition of unity ${f_i}$ 
subordinated to a chosen covering. By (\ref{ginv}) each of the terms appearing 
will be invariant w.r.t. the structure group. So the last 
formula is valid over M understanding each ${\cal L}$ as $\sum_i f_i {\cal 
L}$. $\Box$
\end{lemma}

We began with a smooth manifold with an Euler Density action
which is completely independent of the choice of $\omega_0$. This
gives only a topological invariant of the manifold and is entirely
independent of any embedded hypersurfaces. The $\omega_i$'s, as
well as their number, are arbitrary also. So we see that we have 
constructed a `theory of gravity', in the presence of arbitrarily intersecting
hypersurfaces of discontinuity in the connection, which is a topological invariant. 
It is a trivial theory in that the action is completely insensitive to these 
hypersurfaces. The `gravitational' equations of motion vanish identically, regardless
of the geometry, providing no way to relate geometry to energy-momentum.

\subsubsection{Dimensionally continued Euler densities}

Now we consider the dimensionally continued Euler density
for arbitrarily intersecting hypersurfaces separating bulk regions
counted by $i$. We postulate the action:
\begin{gather}\label{fundaction}
S_g=\sum_i \int_{i} {\cal L}_g(\omega_i,e) + \sum^h_{k=2}
\frac{1}{k!} \sum_{i_1..i_k} \int_{\{i_1...i_k\}} {\cal
L}_g(\omega_{i_1},..,\omega_{i_k},e)
\end{gather}


We will show that this action is `one and a half order' in the connection.  
We will need to
revisit our derivation of the composition rule in section
\ref{biat}, this time interpolating between the different metric
functions $E^i(x)$, where the index represents the region (the
local Lorentz index being suppressed). Physically, we require the
metric being continuous at a surface $\Sigma_{1...p+1}$: $i^*E^i
=E$ which implies $i^*(e^i)=e$. Here $i^*$ is the pullback of the
embedding of $\Sigma_{1...p+1}$ into M. We will see that this
continuity condition arises naturally from the action principle.
Define the Lagrangian on the surface $\Sigma_{1...p+1}$ to be:
\begin{align}\label{palatini}
{\cal L}(\omega^1,..,\omega^{p+1},e) = & \int_0^1 dt_1..dt_{p} \
\zeta_{p}f(\theta^1,\theta^2,..\theta^{p},\Omega_{p},..\Omega_{p}
,e_{p}),
\\\nonumber
(e_p)_{a_1...a_{2n}} = & \frac{1}{(d-2n)!}
(E_{p})^{a_{2n+1}}\wedge..\wedge(E_{p})^{a_{d}\
}\epsilon_{a_1...a_d}.
\end{align}
where
$E_{p}=E^1-(1-t_1)(E^1-E^2)-...-(1-t_1)...(1-t_{p})(E^{p}-E^{p+1})$
and $\zeta_p$ is given by (\ref{zeta}).

Following through the calculation of section (\ref{biat}) , we
pick up extra terms, involving derivatives of $E_{p-1}$, from
using Leibnitz Rule on f.
\begin{eqnarray*}
&& \frac{\partial}{\partial t_{p+1}}
f(\theta^1_{p+1},\theta^2_{p+1},..\theta^p_{p+1},\Omega_{p+1},..\Omega_{p+1},e_{p+1})=
\\ &&
\sum_{s=1}^{p+1} f(\theta^1_{p+1},..,\widehat{\theta^s_{p+1}}
,..,\theta^{p+1}_{p+1},\Omega_{p+1},..\Omega_{p+1}, \frac{\partial
e_{p+1}}{\partial t_{s}}) +\\&& (n-p)
f(\theta^1_{p+1},\theta^2_{p+1},.\theta^{p+1}_{p+1},
\Omega_{p+1},..\Omega_{p+1},D_{p+1}e_{p+1})+(...). \nonumber
\end{eqnarray*}
The (...) are terms which appear just as in section (\ref{biat}).

We will verify our assertion that the action is one-and-a-half 
order by infinitesimally varying the metric and connection in one region. We
vary them as independent fields. Using $t_{p+1}$ to interpolate
between $E^{p+1}$ and $E^{p+1}+\delta E^{p+1}$ and the
corresponding variation of $\omega^{p+1}$:
\begin{gather}
\delta{\cal L} (\omega^1,..,{\omega^{p+1}},e) =  \int_0^1
dt_1..dt_{p+1} \ \zeta_p\ \Xi +(...),
\end{gather}
\begin{align}\label{Xi}
\Xi = & \prod_{i=1}^p (1-t_i) \sum_{s=1}^{p}
f(\theta^1_{p+1},..,\widehat{\theta^s_{p+1}}
,..,\delta\omega^{p+1},\Omega_{p+1},..\Omega_{p+1}, \frac{\partial
e_{p+1}}{\partial t_{s}}) \\\nonumber & -
f(\theta^1_{p+1},..,\theta^{p}_{p+1},\Omega_{p+1},..\Omega_{p+1},
\frac{\partial e_{p+1}}{\partial t_{p+1}})\\\nonumber
& +\prod_{i=1}^p (1-t_i) (n-p+1)
f(\theta^1_{p+1},...\theta^{p}_{p+1},\delta\omega^{p+1},
\Omega_{p+1},..\Omega_{p+1},D_{p+1}e_{p+1}).
\end{align}
The (...) are terms which will cancel when intersections are taken
into account, just as in the topological theory (provided that the
metric is continuous). Above, we have made use of
$\theta^{p+1}_{p+1} =-(1-t_1)...(1-t_p)\delta\omega^{p+1}$.

We require the vanishing of the terms in (\ref{Xi}) involving
$\delta \omega^{p+1}$. Now
$E_{p+1}=E^1-(1-t_1)(E^1-E^2)-...+(1-t_1)...(1-t_{p+1})\delta
E^{p+1}$. Making use of formula (\ref{varvol})
\begin{align}
\frac{\partial }{\partial t_s}(e_{p+1})_{a_1...a_{2n}} = &
\frac{\partial}{\partial t_{s}}(E_{p+1})^{b} \wedge
(e_{p+1})_{a_1...a_{2n}b}\\\nonumber = & \sum_{i=1}^p
(1-t_1)...\widehat{(1-t_s)}...(1-t_i) (E^i-E^{i+1})^b \wedge
(e_p)_{a_1...a_{2n}b}+{\cal O}(\delta E^{p+1}).
\end{align}
So we see the first term in (\ref{Xi}) vanishes if
$i^*(E^{i+1})=i^*(E^i)$ for all $i=1...p+1$ i.e. the metric is
continuous. Given this, we see that:
\begin{align}
i^*(D_{p+1} E_{p+1}) =& \,i^*\left(dE_{p+1} +\omega_{p+1}\wedge
E_{p+1}\right)
\\\nonumber
=&\,i^*\Big(d\big\{E^1+t_1(E^2-E^1)+...t_1...t_{p+1}\delta
E^{p+1}\big\} \Big.\\\nonumber &\qquad+\Big. \big\{\omega_1
+t_1\theta^1+...+t_1...t_p \delta \omega_{p+1}\big\} \wedge (E +
t_1...t_{p+1}\delta E^{p+1})\Big)\\\nonumber &\hspace{-.8in}=
\,i^*\Big(D(\omega^1)E^1+\sum_{i=1}^{p}
t_1...t_i(D(\omega^{i+1})E^{i+1}-D(\omega^i)E^{i}) +{\cal
O}(\delta \omega^{p+1})+{\cal O}(\delta E^{p+1})\Big).
\end{align}
The third term in (\ref{Xi}) already contains a $\delta
\omega^{p+1}$ apart from the $D_{p+1} e_{p+1}$. $D_{p+1}e_{p+1}$
is proportional to $D_{p+1}E_{p+1}$ so to first order in $\delta
E^p$, this term vanishes if $D(\omega^i)E^i = 0$ for all
$i=1...p$. \cite{Enote}

The only non vanishing term in (\ref{Xi}) is the second which
involves:
\begin{align*}
\frac{\partial }{\partial t_{p+1}}(e_{p+1})_{a_1...a_{2n}} = &
-(1-t_1)...(1-t_p)(\delta E^{p+1})^b \wedge
(e_{p+1})_{a_1...a_{2n}b}
\\\nonumber
= & -(\delta e_{p})_{a_1...a_{2n}}
\end{align*}
So we arrive at a simple expression for the variation of the
action, once the equation of motion for the connection
and continuity of the metric have been
substituted.
\begin{gather*}
\delta {\cal L} (\omega^1,...\omega^{p+1},e) = \int_0^1
dt_1...dt_p\  \zeta_p\
f(\theta^1,..,\theta^{p},\Omega_{p},..\Omega_{p}, \delta e)+(...)
\end{gather*}

Then, variation of an $\omega_i$  will
vanish automatically upon imposing the zero torsion condition and the continuity of the metric
at the intersections\cite{Torsionnote}. Second, from the variation of the
frame
$E^a$ we obtain field equation for gravitation and its relation to
the matter present, by
\begin{gather}
\delta_E S_g +\delta_E S_{\text{matter}}=0
\end{gather}
The field equations are actually algebraically obtained, on the
gravity side, using (\ref{varvol}).
Note that although intersections describe physically situation such as
collisions there is a non-zero energy momentum tensor at the
intersection when the theory is not linear in the curvature
2-form. The dimensionally continued n-th Euler density produces a
non-zero energy tensor down to d-n dimensional intersections.
Explicitly, the gravitational equation of motion for a fundamental
intersection $\Sigma_{1..p+1}$, carrying localised matter ${\cal
L}_{m(1...p+1)}$ is
\begin{gather}\label{eqm}
\int_0^1 dt_1...dt_p\  \zeta_p\ \theta^1\wedge..\wedge\theta^{p}
\wedge(\Omega_{p})^{n-p}\wedge\delta E\wedge e =\delta_E {\cal
L}_{m(1...p+1)}.
\end{gather}
We have dropped the local frame index and $(\Omega_p)^{n-p}$ =
$\Omega_p\wedge...\wedge\Omega_p$.

\section{An explicit example}

We calculate the Lagrangian of the simplest
intersection, that of N d-1 dimensional (non-null)
surfaces intersecting at the same d-2
dimensional (non-null) surface.
We then find explicitly the equations of motion for the intersection
in the simplest topological density such that equations of motion
are non trivial, the n=2 Euler density (Gauss-Bonnet term). We
also express the energy conservation in the form of relations
among the energy tensors involved in this case.

\subsection{Equation of motion}

\begin{figure}\label{Ninters}
\begin{center}\mbox{\epsfig{file=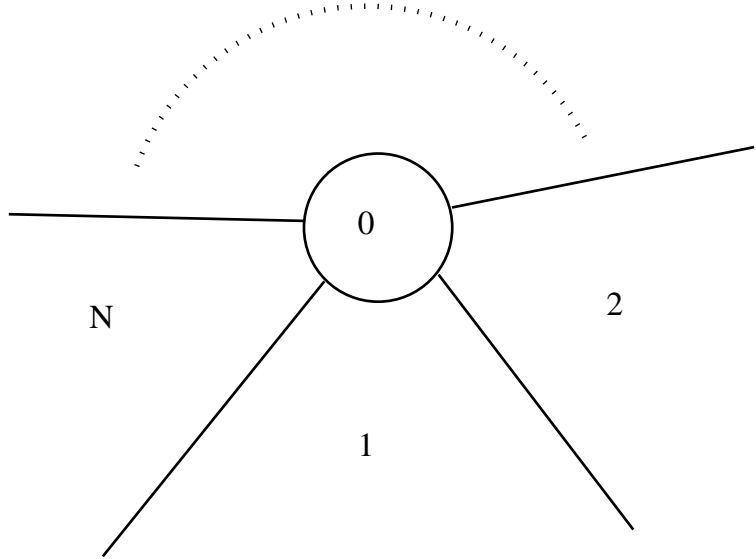, width=10cm}}
\caption{{\small The non-simplicial intersection viewed as the
limit $\{0\}\to I$, where I is a co-dimension 2 surface.}}
\end{center}
\end{figure}

We can treat the non-simplicial intersection with a 2-dimensional
normal space as follows. We divide the space-time
into N+1 regions formed by a N surfaces intersecting a cylinder in
the middle. Taking the cross
section of the system we see a circle with N outgoing lines,
without further intersections (fig. \ref{Ninters}). 
We call $\omega$ the connection
inside the circle and $\omega_i$ the connections of the N
regions formed outside between the lines. We are going to take the
limit of the circle to zero size. The intersections are N lowest
dimensional simplicial intersections. We calculate the contributions at the
intersections implying that they are integrated over the same surface.

The action functional of the hypersurfaces is
\begin{equation}
\int_{12} {\cal L}(\omega_1\omega_2)+\int_{23} {\cal
L}(\omega_2\omega_3)+..+\int_{N1} {\cal L}(\omega_N\omega_1)
\end{equation}
In order to calculate the equation of motion explicitly in terms
of intrinsic and extrinsic curvature tensors we should introduce
the connection $\omega_{ij}$ associated with the induced metric
at the common boundaries. Using the composition rule
\begin{equation} \label{bdycompN}
{\cal L}(\omega_i\omega_j)={\cal L}(\omega_i\omega_{ij})+{\cal
L}(\omega_{ij}\omega_j)- d{\cal L}(\omega_i\omega_j\omega_{ij})
\end{equation}
we obtain one set of contributions at the intersection, when the
common boundary connections are involved.

From the N fundamental intersection, k=3 terms in
(\ref{fundaction}), we have
\begin{equation} \label{Nfundint}
{\cal L}(\omega_1\omega_2\omega)+{\cal
L}(\omega_2\omega_3\omega)+..+{\cal L}(\omega_N\omega_1\omega)
\end{equation}
Up to total derivatives, the expression is
independent of the connection $\omega$, but depends
only on the bulk region connections $\omega_i$. Adding trivially a
set of terms ${\cal L}(\omega_i\omega_1\omega)+{\cal
L}(\omega_1\omega_i\omega)=0, \ i=3..N$ and using the composition
rule we have
\begin{equation} \label{connind}
{\cal L}(\omega_1\omega_2\omega_3)+{\cal
L}(\omega_1\omega_3\omega_4)+..+{\cal
L}(\omega_1\omega_{N-1}\omega_N)
\end{equation}
plus an exact form containing $\omega$. Variation of
(\ref{Nfundint}) with respect to the frame gives us the  equation of motion.

If we want to express things in terms of extrinsic curvatures, we can use
\begin{equation}
{\cal L}(\omega_i\omega_j\omega)+{\cal
L}(\omega_i\omega\omega_{ij})+{\cal L}(\omega\omega_j\omega_{ij})=
{\cal L}(\omega_i\omega_j\omega_{ij})
\end{equation}
(dropping the exact forms integrated on the smooth infinite
intersection) and (\ref{bdycompN}) and (\ref{Nfundint}) to obtain
finally
\begin{eqnarray} \label{Ld-2}
{\cal L}_{d-2}=(12)+ (23)+...+(N1) \ ; \quad (ij)={\cal
L}(\omega_i\omega_{ij}\omega)-{\cal L}(\omega_j\omega_{ij}\omega)
\end{eqnarray}
Clearly $\omega$ can be taken as the connection associated with
the induced metric of the
intersection. Now we can express everything in terms of
the bulk region connections, the second fundamental forms $\theta_{i|ij}$ of 
the surface $\{ij\}$ induced by the region $i$ and the $\chi_{ij}$, the second
fundamental form of the intersection regarded as the boundary of $\{ij\}$.
\begin{equation}
\label{43}
\theta_{i|ij}=\omega_i-\omega_{ij} \ , \qquad
\chi_{ij}=\omega_{ij}-\omega
\end{equation}

Note that, as the form of the Lagrangian
suggests, we could directly try to build the Lagrangian
by applying the method of the last section directly without use of
simplicial intersection and limiting cases. That is, there is
nothing singular in the limit taken.

In order to write the simplest non trivial equation of motion for
the common  intersection of N d-1 dimensional surfaces we consider
the n=2 dimensionally continued Euler density. Applying
(\ref{genl}) or (\ref{genl2}) we find easily
\begin{equation}
{\cal L}(\omega_i\omega_{ij}\omega)=f(\theta_{i|ij},\chi_{ij})
\end{equation}

As noted above, formula (\ref{connind}), the equation of motion for
${\cal L}_{d-2}$ w.r.t. the connection which
remains vanishes via the assumed zero torsion condition in the
dimensionally continued theory.  Varying the frame $E^a$ we
obtain the equations of motion. We define the gravity
Lagrangian as ${\cal L}_g={\cal L}^{(1)}+\alpha_1 {\cal L}^{(2)}$
where ${\cal L}^{(n)}$ is the n-th Euler Density and $\alpha_1$ is
constant of dimension (length)$^2$, the coupling of the
Gauss-Bonnet term.
We express the second fundamental form $\theta^{ab}$ in terms of the 
extrinsic curvature
$K_{ab}$ by
\begin{equation} \label{integrab}
\theta^{ab}=\theta^{cab} E_c \ , \quad \text{where} \quad
\theta^{cab}=-\epsilon(n) 2 \ n^{[a} \nabla^{b]}
n^{c}=-\epsilon(n) 2 \ n^{[a} K^{b]c}
\end{equation}
where $n^{\mu}$ is the normal vector of a (d-1)-dimensional surface embedded 
in a given bulk
and it carries the same indices as the $\theta^{ab}$ (see (\ref{43})) and
$K_{\mu\nu}=h^{\rho}_{\mu} \nabla_{\rho} n_{\nu}$ with 
$h_{\mu\nu}=g_{\mu\nu}-
\epsilon(n) n_{\mu} n_{\nu}, \epsilon(n)=n_{\mu} n^{\mu}=\pm 1$.
The vielbein $e^{\mu}_a$ and its inverse $e^a_{\mu}$ is used to change from 
spacetime to
local frame indices. $\chi^{ab}$  is defined similarly for $v^{\mu}$, the 
normal vector of the
intersection embedded in a given (d-1)-dimensional hypersurface and it 
carries the same indices
as $\chi^{ab}$, and the extrinsic curvature $C_{\mu\nu}=\gamma^{\rho}_{\mu}
h^{\sigma}_{\nu} \nabla_{\rho} v_{\sigma}$ with $\gamma_{\mu\nu}=
h_{\mu\nu}-\epsilon(v) v_{\mu} v_{\nu}$.
We have
\begin{equation} \label{intereqmotion}
2 \alpha_1 \sum \epsilon(n) \epsilon(v)
\left\{ ({\cal K}\bar C)^{ab}+(\bar{{\cal K}}C)^{ab}- \frac{1}{2} \gamma^{ab}
{\rm Tr}({\cal K} \bar C+\bar{{\cal K}}C) \right\} = -T^{ab}_{d-2}
\end{equation}
where ${\cal K}_{ab}$ is the projection of $K_{ab}$ on the intersection. 
Clearly
the sum in (\ref{intereqmotion}) is over all terms in
(\ref{Ld-2}), one for each embedding of each (d-1)-surface in the
adjacent bulk region. We use the notation $\bar{\cal K}_{ab}={\cal
K}_{ab}-\gamma_{ab} {\cal K}$, where ${\cal K}=\gamma^{ab} {\cal
K}_{ab}$, and compact matrix multiplication, for example
$(\bar{\cal K}C)^{ab}=\bar{\cal K}^a_cC^{cb}$. $T^{ab}_{d-2}$ is
the energy momentum tensor which in general should be localised at
the intersection.

\subsection{Energy conservation at the intersection}

Let us see the implications of these results for the question of energy
conservation. We recall that the local expression
of the energy-momentum tensor
conservation is related to the diffeomorphism invariance of the
action, under which the metric tensor changes as $\delta
g_{ab}=2 \nabla_{(a} \xi_{b)}$ where $\xi^a=\delta x^a(x)$ are
infinitesimal coordinate transformations. Note that $2 \nabla_{(a}
\xi_{b)}=\delta g_{ab}$ has to be continuous.

Let us first consider the case of an intersection whose action
term is zero. Let N regions intersect, labelled by $i$, at a
common intersection $I$. We write the energy exchange relations in
the system as
\begin{eqnarray} \label{diffmatter}
\delta_{\xi} S_{\text{matter}}=\sum_i \int_i  T^{ab}_{d} \nabla_a
\xi_{b}+ \frac{1}{2} \sum_{i,j=i\pm1} \int_{ij} T^{ab}_{d-1}
\nabla_a \xi_{b} =0
\end{eqnarray}
where the normal vectors obey $n_{ij}=-n_{ji}$, $j=i\pm1$. Then by
$\xi_b=\xi_{||b}+\epsilon(n) n_b \ n^c \xi_c$ with
$\xi_{||b}=h^c_b \xi_c$ we obtain
\begin{eqnarray} \label{conserv}
&& -\sum_i \int_i   \nabla_a T^{ab}_{d} \xi_b+  \\ &&  \nonumber
+\sum_{ij} \int_{ij} \epsilon(n)  n_a T^{ab}_d h^c_b \ \xi_c-\frac{1}{2}
D_a  T^{ab}_{d-1} \xi_b + \\\nonumber  && +\sum_{ij} \int_{ij} n_a
T^{ab}_d n_b \
n^c \xi_c + \frac{1}{2} T^{ab}_{d-1} K(n)_{ab} \  \epsilon(n)
n^c \xi_c  + \\\nonumber &&+ \int_I  \sum_{ij}
\frac{1}{2}\epsilon(v) v_a T^{ab}_{d-1} \xi_b =0
\end{eqnarray}
where $K(n)_{ab}=h^c_a \nabla_c n_b$ and $n^a, \ K_{ab}$ carry an
index $ij$. Also $j=i\pm1$; the same for $v^a$ which is the normal on $I$
induced by $ij$ pointing outwards. Recall that integrals are taken
over the interiors of the sets. Along with the known relations we
obtain then the ones related with the intersection
\begin{equation} \label{energyedge}
\sum   \epsilon(v)  \ v_a T^{ab}_{d-1} \gamma^c_b =0 \ , \quad
\sum   v_a T^{ab}_{d-1} v_b \ v^c =0
\end{equation}
where the sum is over all shared boundaries. $\gamma_{ab}$ is the
induced metric at $I$.

Equation (\ref{energyedge}) implies that the total
energy current density at the intersection or
collision is zero. This is valid though when the energy
tensor at the intersection vanishes identically.
On the other hand, as we have learned,
the energy tensor is not zero in general and the energy
conservation has to take into account this lower dimensional
energy tensor existing at the intersection hypersurface. In such a
case there is an additional term in
(\ref{diffmatter}) that can be written as
\begin{equation}
\frac{1}{2N} \int_I \sum_{ij} T^{ab}_{d-2} \nabla_a \xi_b
\end{equation}
where we sum over the contribution from each side of every shared
boundary for $N$ regions. $T^{ab}_{d-2}$ is the total energy
momentum tensor on $I$. We decompose $\xi_b=\xi_{||b}+
\epsilon(n) n_b \ n^c \xi_c + \epsilon(v) v_b \ v^c \xi_c$ where
$\xi_{||b}=\gamma_b^c \xi_c$. We then have
\begin{equation}
- \int_I {\cal D}_a T^{ab}_{d-2} \xi_{||b} + \int_I {\cal D}_a (
T^{ab}_{d-2} \xi_{||b} )+ \int_I T^{ab}_{d-2} \frac{1}{N}\sum_{ij}
( \epsilon(n) {\cal K}_{ab} n^c+ \epsilon(v)
C_{ab} v^c)  \xi_c
\end{equation}
${\cal D}$ is the covariant derivative associated with $\gamma$.
The second term is useful when
the intersection is not smooth itself. The energy exchange
relation are then
\begin{eqnarray} \label{energyedge2}
&& \sum  \epsilon(v) \ v_a T^{ab}_{d-1} \gamma^c_b = {\cal D}_a T^{ac}_{d-2} 
\\
&& \sum    v_a T^{ab}_{d-1} v_b \ v^c +T^{ab}_{d-2} \frac{1}{N}
\sum_{ij}  (\epsilon(n) {\cal K}_{ab} n^c+ \epsilon(v)
C_{ab} v^c)=0    \nonumber
\end{eqnarray}
where the first sums are over all shared boundaries.

For a collision of hypersurfaces,
the intersection surface will be space-like. The $v$
vectors are time-like (velocity) vectors.
We assume hypersurface matter of the form:
\begin{gather}
v^av^bT_{ab} = \rho 
\quad \gamma^{a}_{c}T_{ab}v^b =0
\end{gather}
The first of (\ref{energyedge})
is satisfied automatically whilst the second becomes:
\begin{gather}\label{cons}
\sum_{\Lambda} \rho_\Lambda v^a_\Lambda=0
\end{gather}
where the upper case greek index counts the hypersurfaces.
We can recover the results of Langlois, Maeda and 
Wands\cite{Langlois:2001uq} by first introducing the ortho-normal 
basis at the intersection. The basis is taken to lign up with 
the two vectors $v_\Lambda$ and $n_\Lambda$ of one of the hypersurfaces.
\begin{gather}
E_{(0)}=v_\Lambda, \quad E_{(1)} = n_\Lambda
\end{gather}
We can write the other $v$ vectors in the following
way, motivated by special relativity,
\begin{gather}
v_\Xi = \gamma_{\Xi |\Lambda} E_{(0)} + \gamma_{\Xi |\Lambda}
\beta_{\Xi |\Lambda} E_{(1)}
\end{gather}
where the $\beta$ and $\gamma$ have the usual interpretation
from S.R.
Hence, the two components of equation (\ref{cons}) are:
\begin{align}
\sum_\Xi {\rho_\Xi}\gamma_{\Xi|\Lambda} = & 0\\
\sum_\Xi {\rho_\Xi}\gamma_{\Xi|\Lambda}\beta_{\Xi|\Lambda} = & 0
\end{align}
These are the results found in \cite{Langlois:2001uq}. They are conservation
of energy and momentum respectively.

The hypersurfaces obey the same rules in terms of the
local inertial frame as do point particle collisions in two dimensions. 
This is true for quite general 
bulk backgrounds. The only essential feature is the 
absence of a deficit angle at the collision. This means that there
is a well defined local inertial frame at the collision
and the S.R. addition of velocities applies.

We have calculated the contribution to the energy-momentum
tensor at the collision due to the junction conditions. Our calculation
implicitly assumed that there was no conical singularity (see footnote to 
Lemma \ref{lem3}).
There may be some correction to this from a conical singularity.
If we impose some reasonable energy condition such as the dominant 
energy condition, this space-like matter should vanish- the 
two contributions should cancel. The assumption of no conical 
deficit is then justified 
for the {\it Einstein} theory, because we have seen that there
is no contribution due to the junction conditions.
But this would not be so for the Gauss-Bonnet theory. In that case, 
the cancellation would demand that there be a conical singularity at 
the collision.
Conversely, if we impose that there be no such singularity,
we must have space-like matter localised at the collision.
\\\\

Since completion of this work, 
intersecting branes in Lovelock gravity have been studied
by Lee and Tasinato\cite{Tasinato}
and by Navarro and Santiago \cite{Navarro}.
Further analysis of the geometry of intersections
has been done by us in \cite{Gravanis-04}.
\\\\

$\bf{ \qquad Acknowledgements}$

The work of EG was supported by a King's College Research Scholarship (KRS).
The work of SW was supported by EPSRC.


\begin{thebibliography}{99}
\bibitem{Pontryagin_note}
If the Lagrangian is taken to be a sum of such terms 
there are solutions of non-zero torsion. This makes it natural to consider 
also dimensionally continued Pontryagin densities which contain torsion
explicitly, see $\cite{Mardones}$.
\bibitem{CSboundary}
In different contexts the boundary variation of
Chern-Simons terms are important
(e.g. Gravitational and gauge anomalies$\cite{Ginsparg}$
and black hole thermodynamics$\cite{Gegensberg}$).
Here ${\cal L}_3$ is the boundary action of three
`intersecting' Chern-Simons theories and as such it is
gauge invariant under local Lorentz transformations.
\bibitem{Enote}
The general gravitation theory we consider is
built as a sum of a desired set of dimensionally continued
topological densities. The coefficients can be taken to be
functions of scalar fields, making the theory dilatonic. The
variation with respect to the connection leads to a non-zero
torsion in this case. The torsion equation is a constraint on the
variation of the scalar fields. Solving it for the connection and
substituting in the Lagrangian one obtains explicitly an action
for the dilatonic fields.
\bibitem{Torsionnote}
The equations of motion
for $\omega$ are satisfied by the zero torsion condition
but are not, for $n \geq 1$ identical to it.
There are potentially solutions
of non-vanishing torsion, see \cite{Mardones}.
\end{thebibliography}
  \end{document}